# POPULATION SYNTHESIS OF X-RAY SOURCES AT THE GALACTIC CENTER

S.B. Popov[1](ps@sai.msu.su), V.M. Lipunov[1], L.M. Ozernoy[2], K.A. Postnov[1], M.E. Prokhorov[1]

[1]Sternberg Astronomical Institute, Universitetskij pr. 13, 119899 Moscow

[2]Physics Dept. and Inst. for Comput. Sci. & Inform., George Mason Univ.,
also Lab. for Astron. & Solar Phys., NASA/GSFC

## Abstract

We are exploring the evolution of massive binary star populations for a nuclear starburst occuring in the conditions similar to the Milky Way galaxy, in its central few pc, on a time scale of 10 Myr. A computer code is applied allowing for computing, by using Monte Carlo simulations, the evolution of a large ensemble of binary systems, with proper accounting for spin evolution of magnetized neutron stars (NSs). Our results include the number of X-ray transients (NS + main sequence star), super-accreting black holes (BHs), and binaries consisting of a BH + supergiant, all as functions of time.

We find that by 5 Myr after such a starburst one expects $\approx$ 1 X-ray sources with a BH (Cyg X-1 type), $\approx$ 1 SS 433-like source and $\approx$ 54 transient sources with a NS, all to be within the central kpc. An interesting result that can be considered as a specific starburst feature, is that the ratio of the number of systems like SS 433 to the number of X-ray transients is about 1:100, compared to 1:1000, characteristic to the average ratio in the galactic field. The ratio of the number of X-ray sources containing a BH to the number of X-ray transients with NSs turns out to be a sensitive function of the age of the starburst, and its computed value $\approx$ 0.03 is consistent with observations.

# 1 Introduction

X-ray observations of the Galactic center have revealed a number of energetic X-ray sources to be located in the innermost regions of the Galaxy (Churasov et al[1]). Here, we apply Monte Carlo simulations to the evolution of a large ensemble of binary systems, with proper accounting for spin evolution of magnetized compact stars such as white dwarfs (WDs) and neutron stars (NSs) (Kornilov & Lipunov [2], Lipunov[3], Lipunov et al [4], Lipunov [5], Lipunov et al.[6] ). In this poster, we will focus on the most prominent, from the observational point of view, representatives of the late stages of massive binary evolution, such as X-ray transients (NS in a highly eccentric orbit around a main sequence star, like A0535+26), super-accreting BHs (observationally seen as SS 433 if in pair with a Roche lobe filling secondary component), and BH + supergiant binaries (like Cyg X-1).

# 2 The model

1. We consider only massive binaries, i.e. those having the mass of the primary (more massive) component in the range of $10 - 120$ $M_\odot$. Only they can evolve off the main sequence during the time as short as 10 Myr to yield compact remnants (NSs and BHs).
2. Initial primary components' masses are distributed according to the Salpeter's mass function with an power law exponent $\alpha = 2.35$.
3. The initial mass ratio distribution was taken in a power law form $f(q) \propto q^2$ .
4. Initial eccentricity is assumed to be zero.
5. After the collaps an additional kick velocity, $v_{kick}$, should be imparted to the newborn NS. In the present calculations, the kick velocity was taken to be 75 km/s.
6. We consider stars with a constant (solar) chemical composition.
7. The process of mass transfer between the binary components is treated, when appropriate, as a conservative one, i.e. the total angular momentum of the binary system is considered to be constant.
8. To get statistically significant results, the evoluiton of 300,000 binary systems have been computed. Then we normalized the figures so as to be in agreement with the Tamblyn & Rieke[9] calculations of the number of massive OB-stars that survive $\sim 7$ Myr after the starburst.

# 3   The results

The evolution of some selected types of X-ray binaries during the first 10 million years after the onset of star formation burst is presented in Figure 1.

First we show (Fig. 1a) the number of X-ray transient sources consisting of a NS in an eccentric orbit around a massive secondary that acquired enough angular momentum during the mass exchange stage to become a rapidly rotating Be star.

Then we show the evolution of a BH containing X-ray binaries of Cyg X-1 type (Fig. 1b) and of those with a superaccreting BH of SS 433 type (Fig. 1c). We note that the well known X-ray source 1E 1740-2942 (Churasov et al[1]) may belong to SS 433-like binaries.

One of the most interesting results concerns the relative number of SS 433-like binaries together with Cyg X-1–like binaries and X-ray transient sources. Our calculations yield for the ratio of these classes of sources to be $\sim 1/50$, which significantly exceeds this ratio for the entire Galaxy.

The number ratio of a BH containing binaries (of both SS 433 and Cyg X-1 type) to the X-ray transients with Be-stars is plotted in Fig. 1d. It can be seen that this ratio is very sensitive to the time elapsed after the starburst and therefore it can serve as a tool to estimate the age of X-ray binaries at the Galactic center.

In addition, the calculated a simple model spatial distribution of these sources (Fig.2).

# 4   Discussion and conclusions

We compare our results with the observational data by using the *GRANAT* X-ray observations of the Galactic center (Churazov et al[1], Pavlinsky et al [8]). 11 X-ray sources have been reported to be observed in the central region of the Galaxy (750 pc × 750 pc), including 2 sources being classified as BH-candidates by their spectra, and 9 being X-ray transients.

As we noted above, the BH-candidates/X-ray transients ratio is a good indicator of the time passed after the onset of the starburst. The computed ratio$\approx 0.2$ at the age of 4 Myr and $\approx 0.03$ at the age of 5 Myr (which should be considered as a lower limit to the true value because we are not able to observe all the X-ray transients simultaneously) roughly corresponds to the observed ratio of ≈ 0.2. Furthermore, in our model at the age of 7 Myr the X-ray transients occupy a region ≈ 2000 pc × 2000 pc in size (but more than the half of them are distributed in the region ≈ 700 pc × 700 pc), as they acquire high velocities (after the faster evolving primary component explodes as a supernova) and move with an initially high velocity dispersion of about 100 km/s. Hence, by considering the central 750 pc × 750 pc region, one will get the ratio closer to the observed value.

To conclude, we note that the modelling of standard binary stellar evolution after a starburst at the Galactic center (Ozernoy[7]) about 4-7 million years ago yields the number of X-ray sources to be consistent with available X-ray observations of the Galactic center.

*Acknowledgements.*
The authors acknowledge I. Panchenko and S. Nazin for useful discussions. The work was partially supported by the COSMION grant and by Grant No JAP100 from the International Science Foundation and Russian Government.

# References

1 Churazov, E. et al. 1994, in "Frontiers of space and ground based astronomy", eds. W.Wamsteker et al., Kluwer Acad. Press
2 Kornilov, V.G., Lipunov, V.M. 1983, AZh, 60, 574
3 Lipunov, V.M. 1992, Astrophysics of Neutron Stars, Springer Verlag, Heidelberg
4 Lipunov, V.M., Nazin, S.N., Panchenko, I.E., Postnov, K.A. & Prokhorov, M.E. 1995, A & A 298, 677
5 Lipunov, V.M. 1994, in "Mem. Soc. Astr. It.", V.65, p.21
6 Lipunov, V.M., Postnov, K.A., Prokhorov, M.E. & Osminkin, E.Yu. 1994, ApJ 423, L121
7 Ozernoy, L.M. 1994, in "Multi-Wavelength Continuum Emission of AGN", eds. T.J.-L. Courvoisier & A. Blecha, Kluwer, p. 351
8 Pavlinsky, M.N., Grebenev, S.A., and Sunyaev, R.A. 1994, ApJ 425, 110
9 Tamblyn, P. & Rieke, G.H. 1993, ApJ 414, 573

Table 1: Evolution of the selected types of X-ray sources

| Age | BH/NS | SS433 | Cyg X-1 | Transients |
|-----|-------|-------|---------|------------|
| 2.0 | 1.53E+02 | 2.87E+01 | 2.25E+01 | 3.33E-01 |
| 2.5 | 1.55E-00 | 3.02E+00 | 1.17E+01 | 9.47E+00 |
| 3.0 | 4.54E-01 | 1.25E+00 | 7.86E+00 | 2.01E+01 |
| 3.5 | 1.16E-01 | 8.25E-01 | 2.56E+00 | 2.91E+01 |
| 4.0 | 6.74E-02 | 6.84E-01 | 1.91E+00 | 3.85E+01 |
| 4.5 | 3.69E-02 | 5.79E-01 | 1.21E+00 | 4.85E+01 |
| 5.0 | 2.64E-02 | 5.61E-01 | 8.77E-01 | 5.45E+01 |
| 5.5 | 1.54E-02 | 4.91E-01 | 4.91E-01 | 6.39E+01 |
| 6.0 | 8.91E-03 | 3.33E-01 | 2.81E-01 | 6.89E+01 |
| 6.5 | 1.08E-02 | 4.04E-01 | 4.04E-01 | 7.45E+01 |
| 7.0 | 7.18E-03 | 4.56E-01 | 1.23E-01 | 8.06E+01 |
| 7.5 | 5.59E-03 | 3.16E-01 | 1.58E-01 | 8.47E+01 |
| 8.0 | 6.46E-03 | 3.33E-01 | 2.28E-01 | 8.68E+01 |
| 8.5 | 5.63E-03 | 3.51E-01 | 1.58E-01 | 9.04E+01 |
| 9.0 | 5.84E-03 | 3.86E-01 | 1.58E-01 | 9.31E+01 |
| 9.5 | 4.72E-03 | 3.51E-01 | 8.77E-02 | 9.29E+01 |

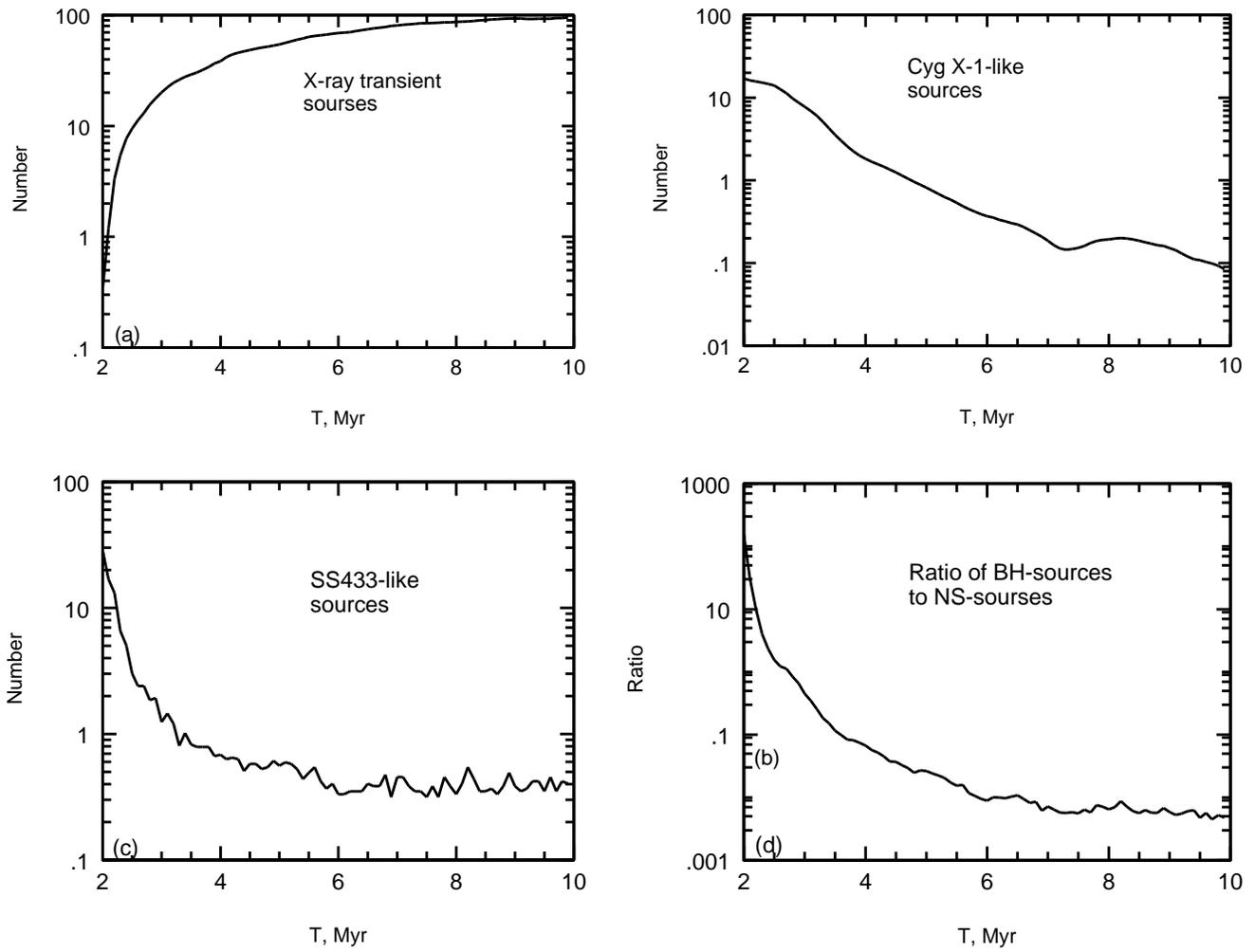

Figure 1: The evolution of some selected types of X-ray binaries during the starburst at the Galactic center.

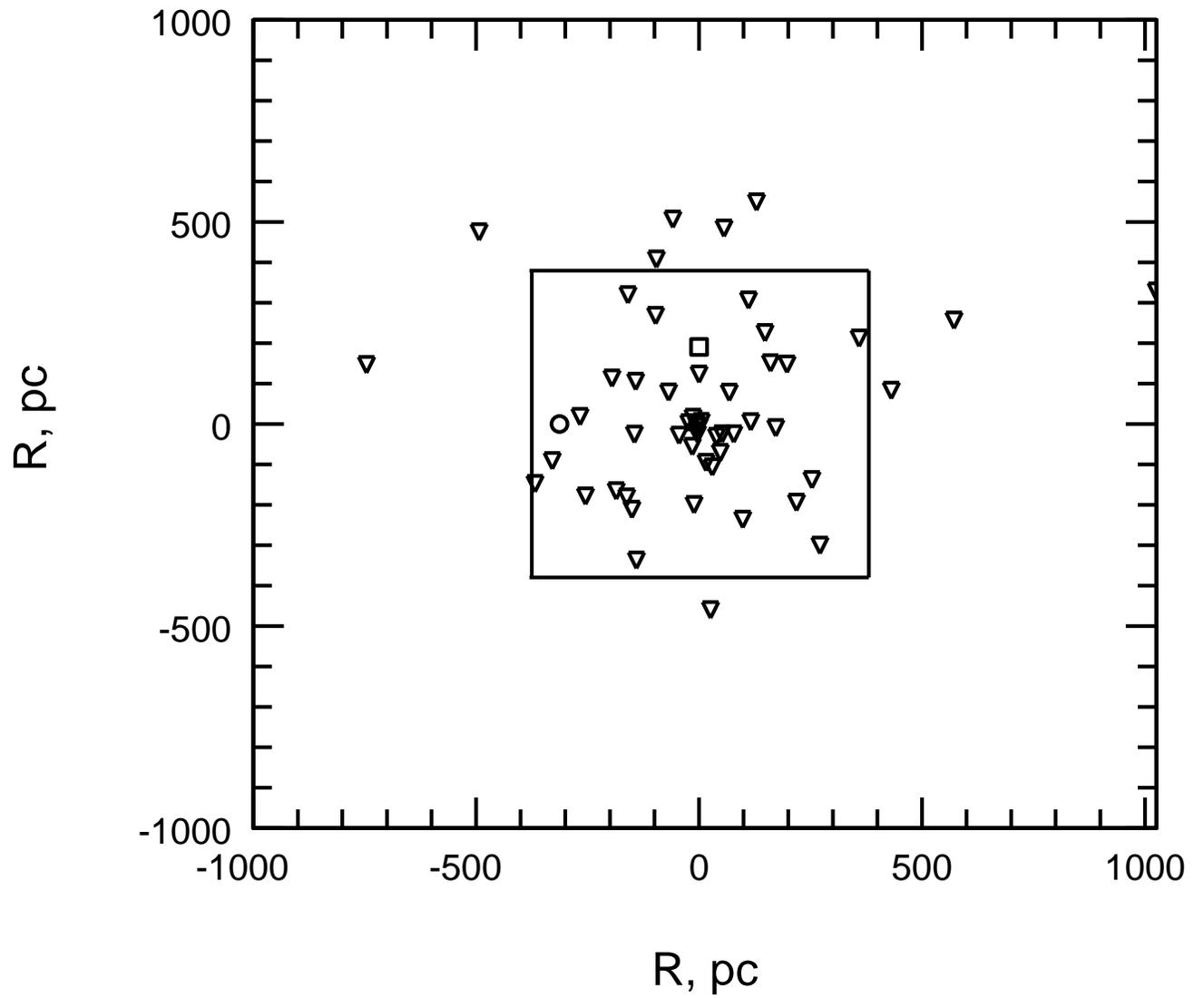

Figure 2: 2-D projection of the spatial distribution of some selected types of X-ray binaries 5 Myr after the onset of the starburst at the Galactic center. Triangles– X-ray transients, square– SS433-like sorce, circle– Cyg X-1-like source.

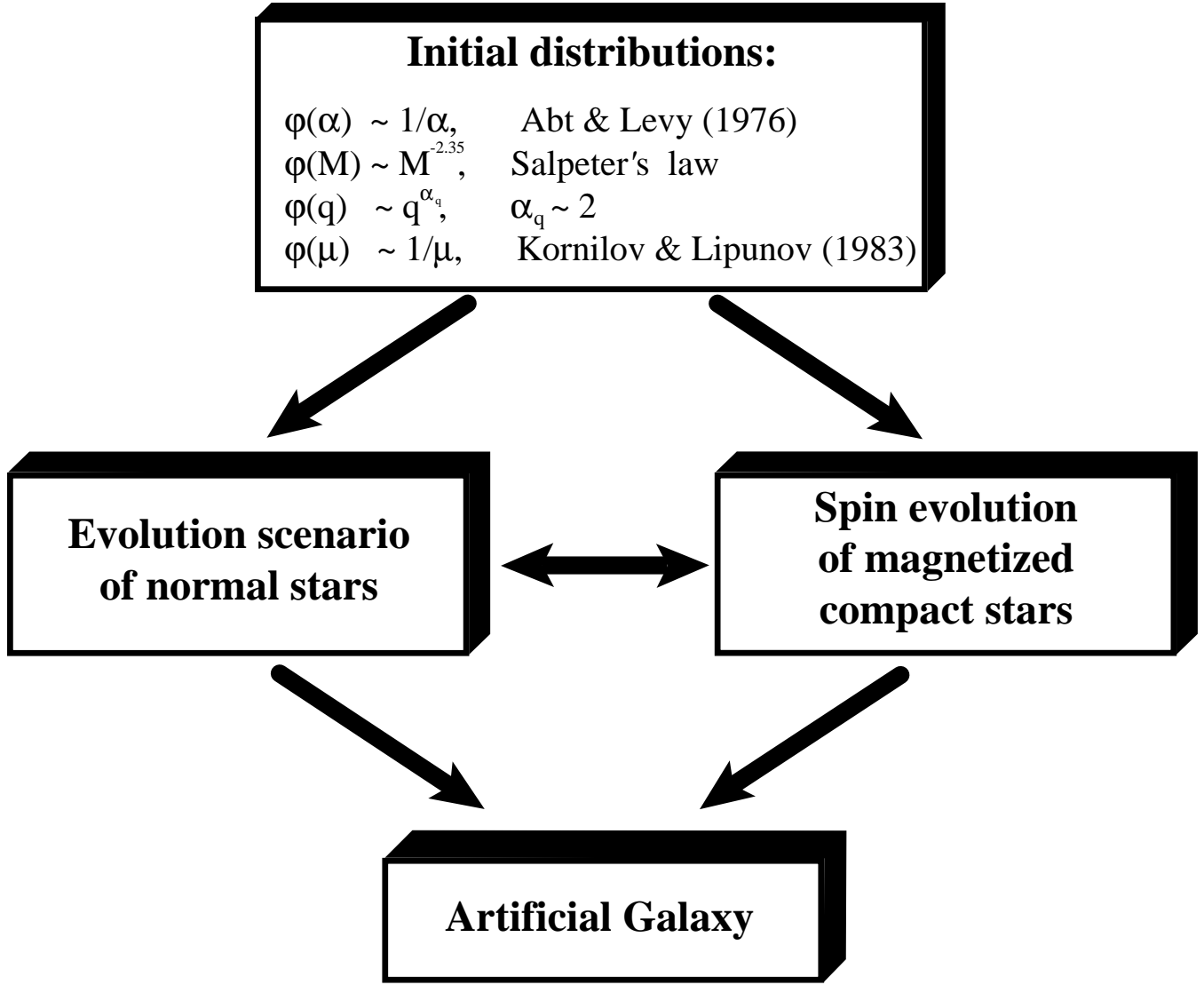

Figure 3: Parameters of the model.